# Superconductivity in Layered Pnictides $BaRh_2P_2$ and $BaIr_2P_2$

*Daigorou HIRA*[1,2], Tomohiro TAKAYAMA[1,2], Ryuji HIGASHINAKA[3],*

*Hiroko ARUGA-KATORI[3], Hidenori TAKAGI [1,2,3]*

[1] *Department of Advanced Materials, University of Tokyo, Kashiwa, Chiba 277-8561*
[2] *JST, TRIP, Kashiwa, Chiba 277-8561*
[3] *RIKEN (The institute of Physical and Chemical Research), Wako, Saitama 351-0198*

Bulk superconductivity was discovered in $BaRh_2P_2$ ($T_c$ = 1.0 K) and $BaIr_2P_2$ ($T_c$ = 2.1 K), which are isostructural to $(Ba,K)Fe_2As_2$, indicative of the appearance of superconductivity over a wide variety of layered transition metal pnictides. The electronic specific heat coefficient $\gamma$ in the normal state, 9.75 and 6.86 mJ/mol $K^2$ for $BaRh_2P_2$ and $BaIr_2P_2$ respectively, indicate that the electronic density of states of these two compounds are moderately large but smaller than those of Fe pnictide superconductors. The Wilson ration $R_W = \chi / \gamma \times \pi^2 k_B^2 / 3\mu_B^2$ close to 1 indeed implies the absence of strong electron correlations and magnetic fluctuations unlike Fe pnictides.

Recent discovery of superconductivity in Fe pnictides has provided fresh impetus to the exploration of novel superconductors with a high transition temperature. After the first report on superconductivity in $LaFeAs(O,F)$,[1] the maximum superconducting transition temperature $T_c$ of Fe pnictides quickly reached 55 K for $SmFeAs(O,F)$[2] by replacing La with the other rare earth elements. Further work has revealed that not only $REFeAs(O,F)$ with $ZrCuSiAs$-type structure but also a variety of iron-based compounds with squared lattices of tetrahedrally coordinated Fe show superconductivity at relatively high temperatures, including $(Ba,K)Fe_2As_2$ ($T_c$ = 38 K)[3] with $ThCr_2As_2$-type structure, $LiFeAs$ ($T_c$ = 18 K)[4] with $CuSb_2$-type structure and $FeSe$ ($T_c$ = 8 K)[5] with α-PbO-type structure.

The exploration of new superconductors, triggered by the discovery of $LaFeAs(O,F)$, has concentrated mostly on Fe-based pnictides. A variety of non-Fe pnictides, isostructural to Fe pnictide superconductors, have been known for a long time but not yet fully explored in terms of possible superconductivity. The limited number of non Fe pnictide superconductors known to date, include $LaNiPnO$ : $Pn$ = P, As, ($T_c$ ~ 3 K),[6-8] $ANi_2Pn_2$ : $Pn$ = P, As and $A$ = Sr, Ba ($T_c$ ~ 3 K)[9-11] and $LaRu_2P_2$ ($T_c$ = 4.1 K).[12] Further exploration of non-Fe pnitide superconductors is important for understanding the key factors in realizing the high transition temperature in the Fe pnictides and, if $T_c$ is reasonably high, it will enhance the potential of pnictides substantially as un-tapped reservoir for discovering new superconductors.

In this work, we present new Ir and Rh pnictide superconductors, isostructural to $BaFe_2As_2$, $BaRh_2P_2$ and $BaIr_2P_2$, with $T_c$ = 1.0 and 2.1 K respectively. The existences of $BaRh_2P_2$ and $BaIr_2P_2$ have been known since the early report by Wurth *et al*.[13] and Löhken *et al*.[14] These discoveries demonstrate that a wide variety of pnictides with various transition metal elements can show superconductivity.

The polycrystalline samples used in this study were synthesized by a conventional solid state reaction. Ba, Rh(Ir) and P powders were mixed at a molar ratio of 1:2:2 in argon-filled dry box. The mixture were heated in an evacuated silica tube initially at 400 ˚C for 16 h and then at 1000 ˚C for 16 h. The products, colored dark gray, were chemically stable in air. The obtained samples were characterized by powder X-ray diffraction (XRD Rigaku; RINT Ultima III) with Cu K$\alpha$ radiation. Magnetic, transport and thermal measurements were conducted by using magnetic property measurement system (MPMS; Quantum Design), physical property measurement system (PPMS; Quantum Design), heat capacity installed in a in $^3$He refrigerator (Heliox, Oxford) and home made AC magnetometer installed in a dilution refrigerator (Kelvinox, Oxford).

All the X-ray diffraction peaks of $BaRh_2P_2$ and $BaIr_2P_2$ could be indexed by a tetragonal cell with lattice constants $a$ = 3.939(1) Å, $c$ = 12.576(2) Å and $a$ = 3.946(1) Å, $c$ = 12.572(2) Å, respectively, except for one extra tiny peak in $BaRh_2P_2$ likely originating from impurities (Fig. 1). This confirms that the prepared samples are almost single phase. The XRD pattern was refined reasonably well with a space group $I4/mmm$, indicating that $BaRh_2P_2$ and $BaIr_2P_2$ crystallize in $ThCr_2Si_2$-type structure. The lattice constants are larger than those of $BaFe_2P_2$[12,15] ($a$ = 3.840(1) Å, $c$ = 12.442(1) Å), very likely representing the large ionic radius of 4$d$ Rh and 5$d$ Ir. Comparing $BaRh_2P_2$ and $BaIr_2P_2$, we notice that the in-plane lattice constant of $BaIr_2P_2$ is longer than that of $BaRh_2P_2$ while the out-of-plane lattice constant is roughly the same. The difference in the in-plane should represent again the difference of ionic radius between 4$d$ Rh and 5$d$ Ir. Combining this in-plane expansion with the almost the same out-of-plane lattice constant, suggests that the $IrP_4$ tetrahedron is more compressed along the $c$-axis than the $RhP_4$ tetrahedron.

* E-mail: dhirai@issp.u-tokyo.ac.jp

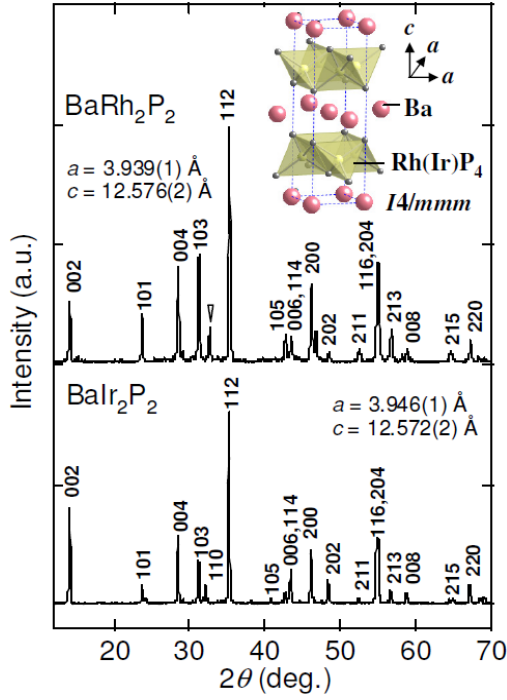

Fig. 1. Powder X-ray diffraction pattern for $BaRh_2P_2$ and $BaIr_2P_2$ collected with Cu Kα radiation at room temperature. All the peaks can be indexed by tetragonal ($I4/mmm$) symmetry and Miller indices are shown for each peak. Inset shows the crystal structure of $BaRh_2P_2$ and $BaIr_2P_2$.

The resistivity and magnetization data indicate that both $BaRh_2P_2$ and $BaIr_2P_2$ are Pauli paramagnetic metals in the normal state. A metallic temperature dependence of resistivity can be clearly recognized for both compounds in Fig. 2 (a). The room temperature resistivity is as large as 10 mΩcm, which though is unreasonably high as a metal, is probably due to grain boundary effects inherent to sintered samples. The magnetic susceptibility in the inset of Fig. 3 shows an almost temperature independent behavior, which can be ascribed the Pauli paramagnetism. The low temperature tail is likely due to the Curie contribution of magnetic impurities. After subtraction of these Curie contributions, the T = 0 K limit Pauli susceptibility $\chi_0$ is estimated to be $1.34 \times 10^{-4}$ and $1.21 \times 10^{-4}$ emu/mol for $BaRh_2P_2$ and $BaIr_2P_2$ respectively. Inspecting the detailed behavior of the susceptibility, weak but clear temperature dependence can be seen only in $BaIr_2P_2$, where the susceptibility increases and eventually saturates with decreasing temperature. This might imply the presence of shallow peak structure in the electronic density of states.

The evidence for superconductivity in $BaRh_2P_2$ and $BaIr_2P_2$ can be found in the resistivity and the magnetization data at low temperatures, shown in Fig.2(b,c) and Fig. 3. At around 1.0 K and 2.1 K for $BaRh_2P_2$ and $BaIr_2P_2$, respectively, a very clear resistance drop to zero resistance state was observed on cooling, accompanied by a large diamagnetic signal indicative of superconductivity. As shown in Fig. 3(b), the zero-field cooling (ZFC) and field cooling (FC) magnetizations at 19.2 Öe for $BaIr_2P_2$ corresponds to 110 and 50% of the perfect diamagnetism, respectively, evidencing that the superconductivity occurs in bulk. Further support for the bulk superconductivity was obtained from the temperature dependent specific heat $C(T)$ shown in Fig. 4. At low temperatures, a clear jump of $\Delta C/T_c = 8.4$ and 10.5 mJ/mol K$^2$ is clearly observed at $T_c$, as shown in Fig. 4.

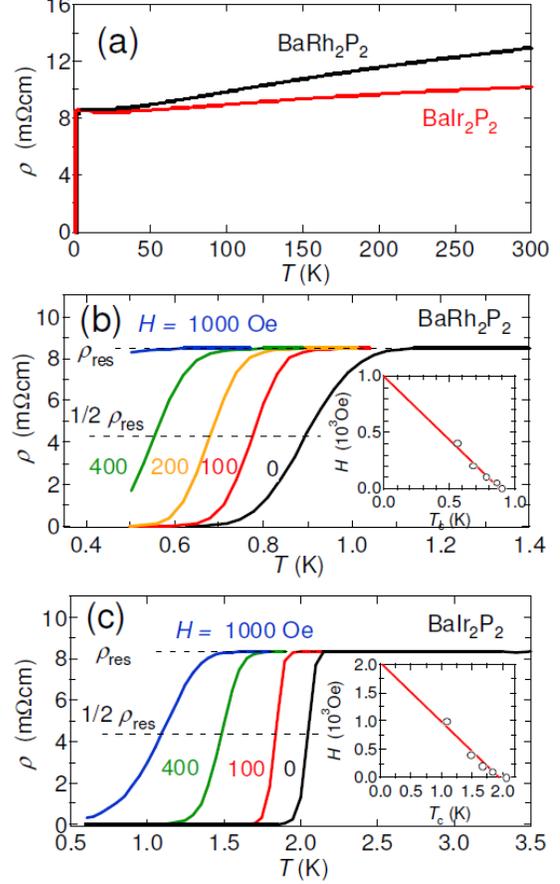

Fig. 2. (a) Temperature dependence of electrical resistivity ρ of $BaRh_2P_2$ and $BaIr_2P_2$ in zero applied field. (b and c) Low temperature resistivity at various applied fields. Inset shows the temperature dependence of the upper critical field ($H_{c2}$) estimated from the midpoint of the superconducting transition ($\rho(T_c) = 1/2 \rho_{res}$).

The upper critical field measurements suggest that these materials are small κ (Ginzbrug-Laudau parameter), type-II superconductors implying that these two systems are relatively clean. The upper critical field $H_{c2}$ was estimated by the field dependence of resistive $T_c$ as shown in the inset of Fig. 2(b) and (c). The initial slope of upper critical field, $dH_{c2}/dT = -1.13$ and $-1.05$ T/K, gives estimate of zero-temperature upper critical fields $H_{c2}(0) = 0.07$ and 0.14 T for $BaRh_2P_2$ and $BaIr_2P_2$, respectively, using the relationship $H_{c2}(0) = -0.7 T_c dH_{c2}/dT_c$.[16] We estimate a Ginzbrug-Landau coherence length $\xi_0 = (\varphi_0/2\pi H_{c2}(0))^{1/2} = 690$ and 480 Å from $H_{c2}(0)$ for $BaRh_2P_2$ and $BaIr_2P_2$, respectively. The low $H_{c2}$ and hence long $\xi_0$, at least in part, should represent the low $T_c$ and the small superconducting gap. Indeed, the estimated values of $\xi_0$ here are comparable to

that of BaNi$_2$As$_2$ ($H_{c2}(0)$ = 0.19 T, $\xi_0$ = 420 Å) with a low $T_c$ = 0.7 K.[12]

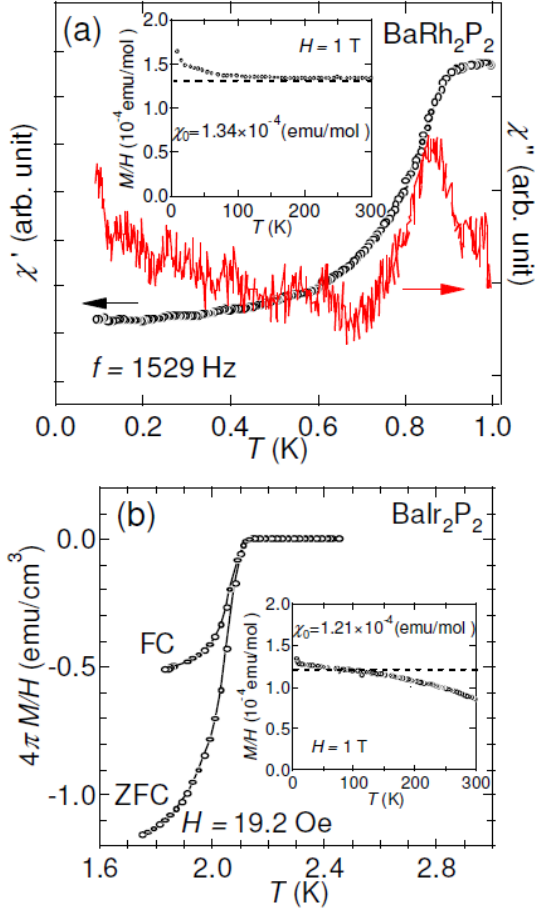

Fig. 3. (a) Temperature dependence of real part (open circle) and imaginary part (solid line) of AC magnetic susceptibility of BaRh$_2$P$_2$ measured with the oscillating frequency of 1529 Hz. (b) Superconducting transition of BaIr$_2$P$_2$ measured in zero-field cooling (ZFC) and field cooling (FC) processes under applied magnetic field of 19.2 Öe. Inset shows magnetic susceptibility ($M/H$) in the wide temperature range. The T = 0 K limit susceptibility $\chi_0$ was estimated by subtracting Curie-like contribution at low temperatures.

The electronic specific heat coefficient $\gamma$ estimated from the specific heat $C(T)$ in the normal state for BaRh$_2$P$_2$ and BaIr$_2$P$_2$ indicates a modest electronic density of state at the Fermi level, not as large as those for FeAs superconductors. The normal-state $C(T)$ from 2 to 5 K can be fitted well to $C(T)/T = \gamma + \beta T^2$ giving $\gamma$ = 9.75 and 6.86 mJ/mol K$^2$ for BaRh$_2$P$_2$ and BaIr$_2$P$_2$, respectively. These $\gamma$ values are typical of transition metal intermetallics but substantially smaller than those reported for high-$T_c$ FeAs superconductors, for example, $\gamma$ = 23 mJ/mol K$^2$ for (Ba,K)Fe$_2$As$_2$.[17] These $\gamma$ values yield a normalized specific heat jump at $T_c$, $\Delta C/\gamma T_c$ = 0.87 and 1.53, which is comparable or smaller than the BCS weak coupling limit value of 1.43. Particularly pronounced deviation from the BCS value in BaRh$_2$P$_2$ may be ascribed to the very broad superconducting transition, which cannot be compensated for by the usual analysis of taking balancing entropy near $T_c$. The coefficient of $T^3$ term, $\beta$, representing the lattice contribution, are estimated to be 0.504 and 0.390 mJ/mol K$^4$, yielding Debye temperatures $\Theta_D$ = 268 and 292 K for BaRh$_2$P$_2$ and BaIr$_2$P$_2$, respectively. This is not very different from other superconductors with ThCr$_2$Si$_2$-type structure, including $\Theta_D$ = 211 K for BaNi$_2$As$_2$ ($T_c$ = 0.7 K)[12] and $\Theta_D$ = 250 K for BaFe$_2$As$_2$ ($T_c$ = 38 K).[16]

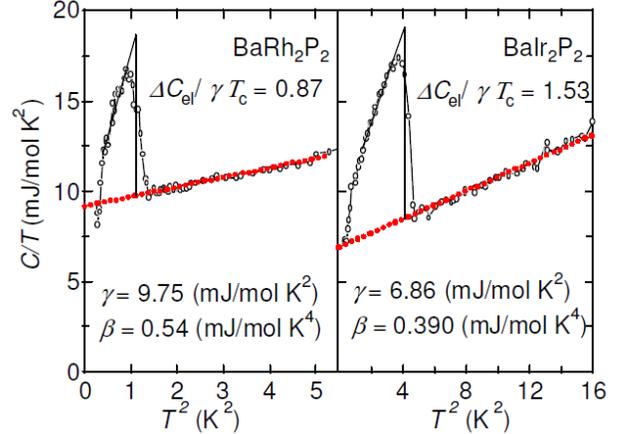

Fig. 4. Specific heat divided by temperature $C(T)/T$ as a function of $T^2$ in zero applied field. The dotted line represents a fit to $C(T)/T = \gamma + \beta T^2$. The entropy conservation construction gives the specific heat jump $\Delta C/\gamma T_c$ = 0.87 and 1.53 for BaRh$_2$P$_2$ and BaIr$_2$P$_2$, respectively

The band calculation of BaRh$_2$P$_2$ and BaIr$_2$P$_2$ performed recently[18] supports the smaller $\gamma$ as compared with FeAs superconductors. It is clear that the reduced density of states comes from two major factors. Firstly, the increase of electron number in the Rh and Ir causes a shift in the Fermi level to a valley region from the peak region where the Fermi level of FeAs superconductors resides. This substantially reduces the DOS. Secondly, 4$d$ and 5$d$ character of conduction band in BaRh$_2$P$_2$ and BaIr$_2$P$_2$ respectively gives rise to an increased band width. These factors might explain, in part, the difference in $T_c$ between the FeAs superconductors, and BaRh$_2$P$_2$ and BaIr$_2$P$_2$

Using the $\gamma$ and $\chi_0$ values obtained experimentally, the Wilson ratio $R_W$ = 0.98 and 1.29 for BaRh$_2$P$_2$ and BaIr$_2$P$_2$, respectively were obtained. These values are close to 1 expected for free electrons. Considering the fact that many strongly correlated Fermi liquids show a Wilson ration close to 2, this implies relatively weak electron correlations both in BaRh$_2$P$_2$ and BaIr$_2$P$_2$. This sharply contrasts with FeAs superconductors where an enhanced Wilson ratio of 11 for LaFeAsO was reported.[19] It is not clear at this stage whether or not this implies the importance of magnetic spin fluctuations.

The comparison between BaRh$_2$P$_2$ and BaIr$_2$P$_2$ might be informative though the difference is much more subtle than the comparison with FeAs superconductors. Despite that the $\gamma$ value of BaIr$_2$P$_2$ is smaller than that of BaRh$_2$P$_2$, the transition temperature $T_c$ of BaIr$_2$P$_2$ is a factor of two higher than BaRh$_2$P$_2$. Note the number of $d$ electrons is exactly the same between the two. It may be interesting to infer the signatures of weak, but visible, correlation effects only in BaIr$_2$P$_2$, including the enhancement of Wilson ratio from 1 and the temperature dependent magnetic susceptibility.

*E-mail: dhirai@issp.u-tokyo.ac.jp

We report superconductivity in both BaIr$_2$P$_2$ and BaRh$_2$P$_2$ with ThCr$_2$Si$_2$-type structure. This discovery demonstrates the presence of superconductivity over a surprisingly broad range of transition metal compounds with ThCr$_2$Si$_2$-type structure from Fe to Ir. This flexibility of superconductivity against transition metal elements provides us with a unique opportunity to explore many new superconductors.

## Acknowledgments

The authors thank M. Nohara, A. Yamamoto, R. S. Perry for stimulating discussion. This work was supported by a Grant-in-Aid for Scientific Research from the Ministry of Education, Culture, Sports, Science and Technology of Japan (Grant No. 19104008 and 16076204) and Global COE Program "the Physical Sciences Frontier", MEXT, Japan.